\documentclass[%twocolumn,
superscriptaddress,aps,preprintnumbers,amsmath,amssymb,prd,nofootinbib,preprint]{revtex4-1}

\usepackage[dvipdfm]{graphicx}
\usepackage{epstopdf}
\usepackage{dcolumn}% Align table columns on decimal point
\usepackage{bm}% bold math
\usepackage{hyperref}
\usepackage{color}

\begin{document}

%%%%%%%%%%%%%%%%%%%%%%%%%%%%%%%%%%%%%%%%%%%

\def\a{\alpha}
\def\b{\beta}
\def\c{\varepsilon}
\def\d{\delta}
\def\e{\epsilon}
\def\f{\phi}
\def\g{\gamma}
\def\h{\theta}
\def\k{\kappa}
\def\l{\lambda}
\def\m{\mu}
\def\n{\nu}
\def\p{\psi}
\def\q{\partial}
\def\r{\rho}
\def\s{\sigma}
\def\t{\tau}
\def\u{\upsilon}
\def\v{\varphi}
\def\w{\omega}
\def\x{\xi}
\def\y{\eta}
\def\z{\zeta}
\def\D{\Delta}
\def\G{\Gamma}
\def\H{\Theta}
\def\L{\Lambda}
\def\F{\Phi}
\def\P{\Psi}
\def\S{\Sigma}

\def\o{\over}
\def\beq{\begin{eqnarray}}
\def\eeq{\end{eqnarray}}
\newcommand{\gsim}{ \mathop{}_{\textstyle \sim}^{\textstyle >} }
\newcommand{\lsim}{ \mathop{}_{\textstyle \sim}^{\textstyle <} }
\newcommand{\vev}[1]{ \left\langle {#1} \right\rangle }
\newcommand{\bra}[1]{ \langle {#1} | }
\newcommand{\ket}[1]{ | {#1} \rangle }
\newcommand{\EV}{ {\rm eV} }
\newcommand{\KEV}{ {\rm keV} }
\newcommand{\MEV}{ {\rm MeV} }
\newcommand{\GEV}{ {\rm GeV} }
\newcommand{\TEV}{ {\rm TeV} }
\newcommand{\1}{\mbox{1}\hspace{-0.25em}\mbox{l}}
\def\diag{\mathop{\rm diag}\nolimits}
\def\Spin{\mathop{\rm Spin}}
\def\SO{\mathop{\rm SO}}
\def\O{\mathop{\rm O}}
\def\SU{\mathop{\rm SU}}
\def\U{\mathop{\rm U}}
\def\Sp{\mathop{\rm Sp}}
\def\SL{\mathop{\rm SL}}
\def\tr{\mathop{\rm tr}}

\def\IJMP{Int.~J.~Mod.~Phys. }
\def\MPL{Mod.~Phys.~Lett. }
\def\NP{Nucl.~Phys. }
\def\PL{Phys.~Lett. }
\def\PR{Phys.~Rev. }
\def\PRL{Phys.~Rev.~Lett. }
\def\PTP{Prog.~Theor.~Phys. }
\def\ZP{Z.~Phys. }

\def\dd{\mathrm{d}}
\def\ff{\mathrm{f}}
\def\BH{{\rm BH}}
\def\inf{{\rm inf}}
\def\ev{{\rm evap}}
\def\eq{{\rm eq}}
\def\SM{{\rm sm}}
\def\Mpl{M_{\rm Pl}}
\def\GeV{{\rm GeV}}
\newcommand{\Red}[1]{\textcolor{red}{#1}}

\def\mDM{m_{\rm DM}}
\def\mphi{m_{\phi}}
\def\TeV{{\rm TeV}}
\def\Gphi{\Gamma_\phi}
\def\TR{T_{\rm RH}}
\def\Br{{\rm Br}}
\def\DM{{\rm DM}}
\def\Eth{E_{\rm th}}
\newcommand{\lmk}{\left(}  
\newcommand{\rmk}{\right)}
\newcommand{\lkk}{\left[}  
\newcommand{\rkk}{\right]}
\newcommand{\lhk}{\left \{ }  
\newcommand{\rhk}{\right \} }
\newcommand{\del}{\partial}  
\newcommand{\la}{\left\langle} 
\newcommand{\ra}{\right\rangle}

%%%%%%%%%%%%%%%%%%%%%%%%%%%%%%%%%%%%%%%%%%%%%%%%%%%%%%%%%%%%%%%

\title{
Discovery of Large Scale Tensor Mode and\\
 Chaotic Inflation in Supergravity
}

\author{Keisuke Harigaya}
\affiliation{Kavli IPMU (WPI), TODIAS, University of Tokyo, Kashiwa, 277-8583, Japan}
\author{Tsutomu T.~Yanagida}
\affiliation{Kavli IPMU (WPI), TODIAS, University of Tokyo, Kashiwa, 277-8583, Japan}
\begin{abstract}
%Recent observations of the cosmic microwave background suggest the existence of tensor mode in large scales, which favors chaotic inflation models.
%Simple chaotic inflation models with power-law potentials are in tension with observations.
%In the context of the supergravity, it has been pointed out that by introducing shift symmetry breaking in the Kahler potential, chaotic inflation models can be consistent with the observation within 1$\sigma$ level.
%However, in the presence of the shift symmetry breaking, it is not clear whether models possess predictability, since higher order terms in the Kahler potential in general affect inflaton dymanics.
%In this letter, we show that chaotic inflation models are consistent with the observation with sufficiently small shift symmetry breaking
%so that higher order terms in the Kahler potential is negligible for the inflaton dynamics and hence model possess predictability.
%Recent observations of the cosmic microwave background suggest the existence of tensor mode in large scales, which favors chaotic inflation models.
The BICEP2 collaboration has recently reported 
%Recent observations of the cosmic microwave background have revealed the existence of
a large tensor fluctuation in the cosmic microwave background, which suggests chaotic inflation models.
In this letter, we reconsider the chaotic inflation model in the supergravity.
We introduce a non-holomorphic shift-symmetry breaking parameter,
which we expect to exist in general,
 and discuss its effect on the inflaton dynamics.
%paying attentions to the effect of higher-dimensional operators.
We show that
%the prediction of
the model predicts a sizable deviation from the original chaotic inflation model and the predicted tensor fluctuation
can lie between the BICEP2 result and the upper bound given by the Planck experiment
with a small shift-symmetry breaking parameter.
%As a result, higher dimensional terms are negligible for the inflaton dynamics and hence
The model is characterized by only two parameters, which yields predictability and testability in future experiments.
\end{abstract}

\date{\today}
\maketitle
\preprint{IPMU 14-0062}
%%%%%%%%%%%%%%%%%%%%%%%%%%%%%%%%%%%%%%%%%%%%%%%%%%%%%%%%%%%%%%%

\section{Introduction}
Cosmic inflation~\cite{Guth:1980zm} is a natural scenario which not only solves the
flatness and the horizon problem,
%and the unwanted relic problem,
but
also explains the large scale structure of the universe and the
fluctuation of the cosmic microwave background (CMB) radiation.
Precise observations
of the CMB~\cite{Hinshaw:2012aka,Ade:2013zuv,Ade:2013uln} begin to reveal nature of inflation.
Recently, the BICEP2 collaboration has reported a large tensor fraction, $r= {\cal O}(0.1)$~\cite{Ade:2014xna},
which favors chaotic inflation models~\cite{Linde:1983gd}.
Chaotic inflation models have been studied in the literature, especially in the context of the supergravity theory (SUGRA).
%in which the theory is well-controlled due to the local supersymmetry.
In this letter, we reconsider chaotic inflation models in the SUGRA.

In SUGRA chaotic inflation models,
%the eta problem~\cite{Ovrut:1983my} is naturally solved by introducing a
the shift-symmetry proposed in Ref.~\cite{Kawasaki:2000yn} is a crucial assumption.
In order to obtain non-zero potential energy, the shift symmetry must be explicitly broken.
In Ref.~\cite{Kawasaki:2000yn} the shift-symmetry breaking is introduced in the superpotential.
%, $W=mX\phi$, where the mass parameter $m$ represents a shift-symmetry breaking.
However, it would be natural to consider that the Kahler potential also has shift-symmetry breaking terms.

The shift-symmetry breaking in the Kahler potential is discussed in Refs.~\cite{Kallosh:2010ug,Li:2013nfa},
and it is shown that the prediction of the model deviates from that of Ref.~\cite{Kawasaki:2000yn} significantly.
%more consistent with observations than models without shift-symmetry breaking.
%There, only single shift symmetry breaking term is considered.
However, it is not clear whether the model possesses predictability.
Higher dimensional terms in the Kahler potential may change inflaton dynamics due to large inflaton field value during inflation,
once the shift-symmetry breaking is introduced.

%In this letter, we show that the chaotic inflation model is consistent with observations with sufficiently small shift symmetry breaking
%so that higher order terms in the Kahler potential is negligible for the inflaton dynamics, and hence the model possesses predictability.
%In the next section, we review the SUGRA chaotic inflation model.

In this letter, we propose to treat the shift-symmetry breaking in a systematic way by introducing a non-holomorphic shift-symmetry breaking spurion ${\cal E}$ and discussing its effect on the inflaton dynamics.
We restrict our attention to the range of the breaking parameter where
higher dimensional terms are negligible for the inflaton dynamics and the model possesses predictability and testability.
% in future experiments.
We show that the prediction for the spectral index and the tensor fraction
can lie between the results of the Planck and the BICEP2 experiments with a small shift-symmetry breaking parameter.
We also show that future observations of the CMB can quantify the reheating temperature of the universe within a factor of ${\cal O}(10)$.
%We show that the chaotic inflation model is consistent with observations with sufficiently small shift symmetry breaking,
%so that higher dimensional terms in the Kahler potential are negligible for the inflaton dynamics, and hence the model possesses predictability and testability.

This letter is organized as follows.
In the next section, we review the SUGRA chaotic inflation model.
In Sec.~\ref{sec:breaking}, we introduce a non-holomorphic shift-symmetry breaking parameter ${\cal E}$ and discuss how the prediction on the spectral index and the tensor fraction is modified by the shift-symmetry breaking.
We estimate the range of the shift symmetry breaking where higher dimensional terms in the Kahler potential are negligible, and show the prediction of the model within the range. 
The last section is devoted to discussion and conclusions.

%There are many possible breaking terms in the Kahler potential. Thus we impose a discrete $Z_2$ symmetry to restrict the shift symmetry breaking term. This discrete symmetry is very important not only to suppress the unwanted linear term is the superpotential, $W=CX$, but also to avoid the gravitino over production\cite{ETY}. In this short letter, we show that a small shift-symmetry breaking parameter, $\epsilon \simeq 0.1$, can explain the observed ........
%Now we see that we need two independent parameters, $m\simeq 10^{-5}$ and  $\epsilon\simeq 10^{-1}$, for the shift-symmetry breaking. The $m$ is  holomorphic and $\epsilon$ non-holomorphic. We hope that a more fundamental theory may explain the origin of such breaking parameters.  

\section{Review on SUGRA chaotic inflation models}
In  this section, we review SUGRA chaotic inflation models in Ref.~\cite{Kawasaki:2000yn}.
For simplicity, we discuss a quadratic chaotic inflation model.
In the SUGRA, the scalar potential is determined by the Kahler potential $K(\phi^i,\phi^{*\bar{i}})$ and the superpotential $W(\phi^i)$,
where $\phi^i$ and $\phi^{*\bar{i}}$ are chiral multiplets and their conjugates, respectively.%
\footnote{We neglect the D-term contribution, which is irrelevant for our purpose.}
The scalar potential is given by
\begin{eqnarray}
\label{eq:potential}
V &=&  e^K \left[
K^{\bar{i}i}D_i W D_{\bar{i}}W^* - 3 |W|^2
\right],\nonumber\\
D_i W &\equiv& W_i + K_i W,
\end{eqnarray}
where subscripts $i$ and $\bar{i}$ denote derivatives with respect to $\phi^{i}$ and $\phi^{*\bar{i}}$, respectively.
$K^{\bar{i}i}$ is the inverse of the matrix $K_{i\bar{i}}$.
Throughout this letter, we use a unit with the reduced Planck mass $M_{\rm pl}\simeq 2.4\times 10^{18}$ GeV being unity.

Chaotic inflation is achieved by introducing two chiral multiplets $\Phi$ and $X$ and assuming the following Kahler and the superpotential,
\begin{eqnarray}
W &=& m X \Phi,\nonumber\\
K &=& K\left(XX^*, \left(\Phi + \Phi^*\right)^2\right) = \frac{1}{2} (\Phi + \Phi^*)^2 + XX^* + \cdots,
\end{eqnarray}
where $\cdots$ denotes higher dimensional terms.
This form of the potentials is realized by assuming an $R$ symmetry, a $Z_2$ symmetry and a shift-symmetry, which are listed in Table~\ref{tab:charge}.
The breaking of the shift symmetry, which is necessary in order to obtain non-zero potential energy, is expressed by the holomorphic spurious field $m$.
The $Z_2$ symmetry is crucial to prevent the over-production of gravitinos in the decay of the inflaton~\cite{Kawasaki:2006gs}.

The inflaton field is identified with the imaginary part of $\Phi$, whose potential from the exponential factor in Eq.~(\ref{eq:potential}) is absent due to the shift symmetry~\cite{Kawasaki:2000yn}, which solves the eta problem~\cite{Ovrut:1983my}. $X$ and the real part of $\Phi$ obtain masses as large as the Hubble scale during inflation by higher dimensional operators and hence are fixed to their origin during inflation. As a result, the potential of the imaginary part of $\Phi$, $\phi$, is given by
\begin{eqnarray}
V (\phi) = \frac{1}{2}m^2 \phi^2,
\end{eqnarray}
which is nothing but the potential of the quadratic chaotic inflation model~\cite{Linde:1983gd}.
The magnitude of the curvature perturbation, ${\cal P}_\zeta\simeq 2.2\times 10^{-9}$, determines the parameter $m$ as (see e.g. Ref.~\cite{Lyth:2007qh})
\begin{eqnarray}
m\simeq 6.0\times 10^{-6} = 1.5\times 10^{13}~{\rm GeV}.
\end{eqnarray}
The spectral index of the curvature perturbation $n_s$ and the tensor fraction $r$ are given by
\begin{eqnarray}
n_s = 1-\frac{2}{N_e} \simeq 0.967~~(N_e = 60),\nonumber\\
r = \frac{8}{N_e} \simeq 0.13~~(N_e = 60),
\end{eqnarray}
where $N_e$ is the number of the e-foldings corresponding to the scale of the interest.
Note that they are determined only by $N_e$, and the model has strong predictability.

\begin{table}
\begin{center}
 \begin{tabular}{|c|c|c|c|c|}
\hline
 & $R$ & $Z_2$ & shift \\
\hline
 $X$ & $2$ & $-1$ & $X\rightarrow X$ \\
 $\Phi$ & $0$ & $-1$ & $\Phi \rightarrow \Phi + i c$\\
$m$ & $0$& $+1$& $m\rightarrow m \frac{\Phi}{\Phi + ic}$\\
\hline
 \end{tabular}
\caption{Charge assignment of (spurious) fields. $c$ is an arbitrary real number.}
\label{tab:charge}
\end{center}
\end{table}

\section{Shift-symmetry breaking in the Kahler potential}
\label{sec:breaking}
In the previous section, we have reviewed the SUGRA chaotic inflation model.
There, we have introduced the shift-symmetry breaking only to the superpotential.
However, it would be more natural to consider that the Kahler potential also has shift-symmetry breaking terms.
The shift-symmetry breaking in the Kahler potential is discussed in Refs.~\cite{Kallosh:2010ug,Li:2013nfa}.
%with only a single shift-symmetry breaking term considered.
However, it is not clear how higher dimensional terms change the prediction of the model
once the shift-symmetry breaking is introduced, since the field value of the inflaton is far above the Planck scale during inflation.

In this section, we propose to treat the shift-symmetry breaking in a systematic way, such that the shift-symmetry breaking is expressed by a non-holomorphic spurious field ${\cal E}$.
We estimate an upper bound on the magnitude of the shift symmetry breaking where higher dimensional terms in the Kahler potential are negligible and hence the model possesses the predictability.
%We show that the model is consistent with observations within 1$\sigma$ level while the upper bound is satisfied.
We restrict our discussion to such a breaking parameter and
show that the prediction of the model can lie between the results of the Planck~\cite{Ade:2013uln} and the BICEP2 experiments~\cite{Ade:2014xna} with a small shift-symmetry breaking parameter.
%As a result,
%so that higher dimensional terms are negligible for the inflaton dynamics.
%and hence the model possesses predictability and testability by future experiments.

%Let us consider in more general frame work, such that the shift-symmetry breaking is expressed by a spurious field ${\cal E}$.
%There are many possible breaking terms in the Kahler potential.
%We parametrize the shift-symmetry breaking in the Kahler potential by a spurious field ${\cal E}$.
The Kahler potential is in general given by
\begin{eqnarray}
K &=& K \left(XX^*,
\left( \Phi + \Phi^*\right)^2,
{\cal E}\left( \Phi - \Phi^*\right)^2\right),
\end{eqnarray}
with the transformation law of ${\cal E}$,
\begin{eqnarray}
{\cal E} \rightarrow {\cal E}\frac{(\Phi-\Phi^*)^2}{(\Phi-\Phi^* + 2ic)^2}.
\end{eqnarray}
As we have mentioned, ${\cal E}$ is non-holomorphic.%
\footnote{
It is possible to introduce the shift symmetry breaking by a holomorphic parameter $m$ as $K\supset (m\Phi-m^*\Phi^*)^2+\cdots$.
However, $m$ is too small to affect the inflaton dynamics.
}
In the following, we estimate the bound on ${\cal E}$ so that ${\cal O} ({\cal E}^2)$ terms do not affect the inflaton dynamics.

The Kahler potential is expanded around the origin as
\begin{eqnarray}
\label{eq:Kahler}
K&=& XX^* + \frac{1}{2}(\Phi +\Phi^*)^2 - \frac{{\cal E}}{2} (\Phi - \Phi^*)^2 + \frac{{\cal E}^2}{4!}\kappa (\Phi - \Phi^*)^4 \cdots,
\end{eqnarray}
where $\cdots$ denotes higher dimensional ${\cal O}({\cal E}^3)$ terms and $\kappa$ is an order one parameter.
The normalization of ${\cal E}$ is fixed by the third term in Eq.~(\ref{eq:Kahler}).
The potential of the inflaton is given by%
\footnote{
The coupling $K \supset  XX^* {\cal E} (\Phi-\Phi^*)^2$ also contributes the scalar potential. The contribution can be absorbed by redefinitions of ${\cal E}$ and $\kappa$ by ${\cal O}(1)$ factors.
}
\begin{eqnarray}
V (\phi) = {\rm exp}\left( {\cal E} \phi^2 + \frac{{\cal E}^2}{6} \kappa\phi^4 + \cdots\right)\times \frac{1}{2}m^2 \phi^2,
\end{eqnarray}
where $\cdots$ denotes higher dimensional ${\cal O}({\cal E}^3)$ terms.%
\footnote{
${\cal O}({\cal E}^2)$ terms in the Kahler potential also contribute to field-dependent kinetic terms of the inflaton field.
%yields the dependence of the wave function on the inflaton field.
In order to simplify the analysis, we neglect them.}
% and estimate the insensitivity to ${\cal O}({\cal E}^2)$ terms by considering the ${\cal O}({\cal E}^2)$ term in the inflaton potential.}
%Here, we have expressed the effect of second lowest dimensional shift-symmetry breaking operators by a parameter $d$, which is of order one.

Let us discuss the dynamics of the inflaton.
The first and the second slow-roll parameter $\epsilon$ and $\eta$ are given by
\begin{eqnarray}
\epsilon(\phi) =& \frac{1}{2}\left(\frac{V_\phi}{V}\right)^2 &\simeq \frac{2}{\phi^2}
\left(
1 + 2{\cal E} \phi^2 + \frac{3+2 \kappa}{3}{\cal E}^2 \phi^4
\right),\nonumber\\
\eta(\phi) =& \frac{V_{\phi\phi}}{V} &\simeq \frac{2}{\phi^2}
\left(
1 + 5{\cal E} \phi^2 + \frac{6+7\kappa}{3} {\cal E}^2 \phi^4
\right),
\end{eqnarray}
where we have neglected ${\cal O}({\cal E}^3)$ terms.
The number of the e-folding $N_e$ and the inflaton field values are related by
\begin{eqnarray}
N_e(\phi) = \int ^\phi_{\phi_{\rm end}}\frac{V}{V_\phi} {\rm d}\phi \simeq 
\frac{1}{4}(\phi^2-\phi^2_{\rm end}) -\frac{{\cal E}}{8}(\phi^4-\phi_{\rm end}^4) + \frac{3-\kappa}{36}{\cal E}^2 (\phi^6 -\phi^6_{\rm end}),
\end{eqnarray}
with $\phi_{\rm end}\simeq \sqrt{2}$.
We have again neglected ${\cal O}({\cal E}^3)$ terms.

\begin{figure}[tb]
\begin{center}
  \includegraphics[width=.6\linewidth]{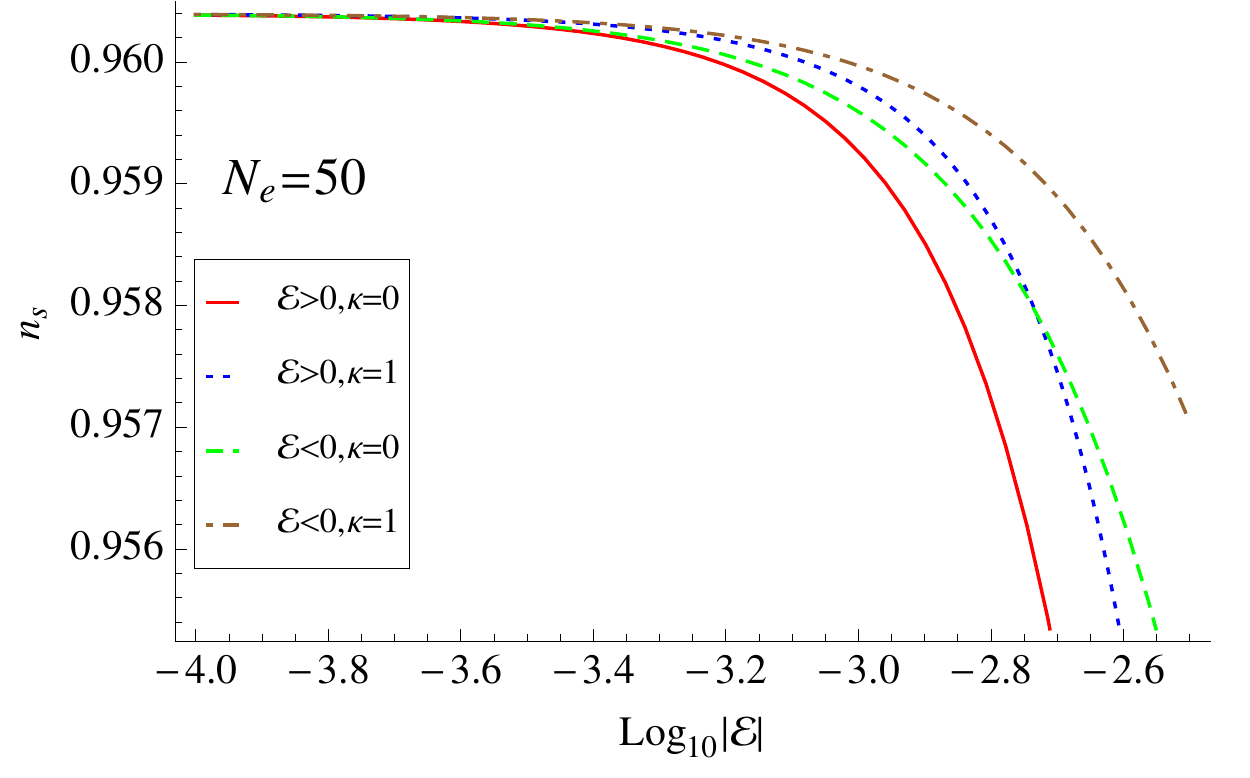}
 \end{center}
\caption{\sl \small
The spectral index $n_s$ as a function of the shift-symmetry breaking parameter ${\cal E}$ for $\kappa=0,1$.
}
\label{fig:tilt}
\end{figure}

\begin{figure}[tb]
\begin{center}
  \includegraphics[width=.6\linewidth]{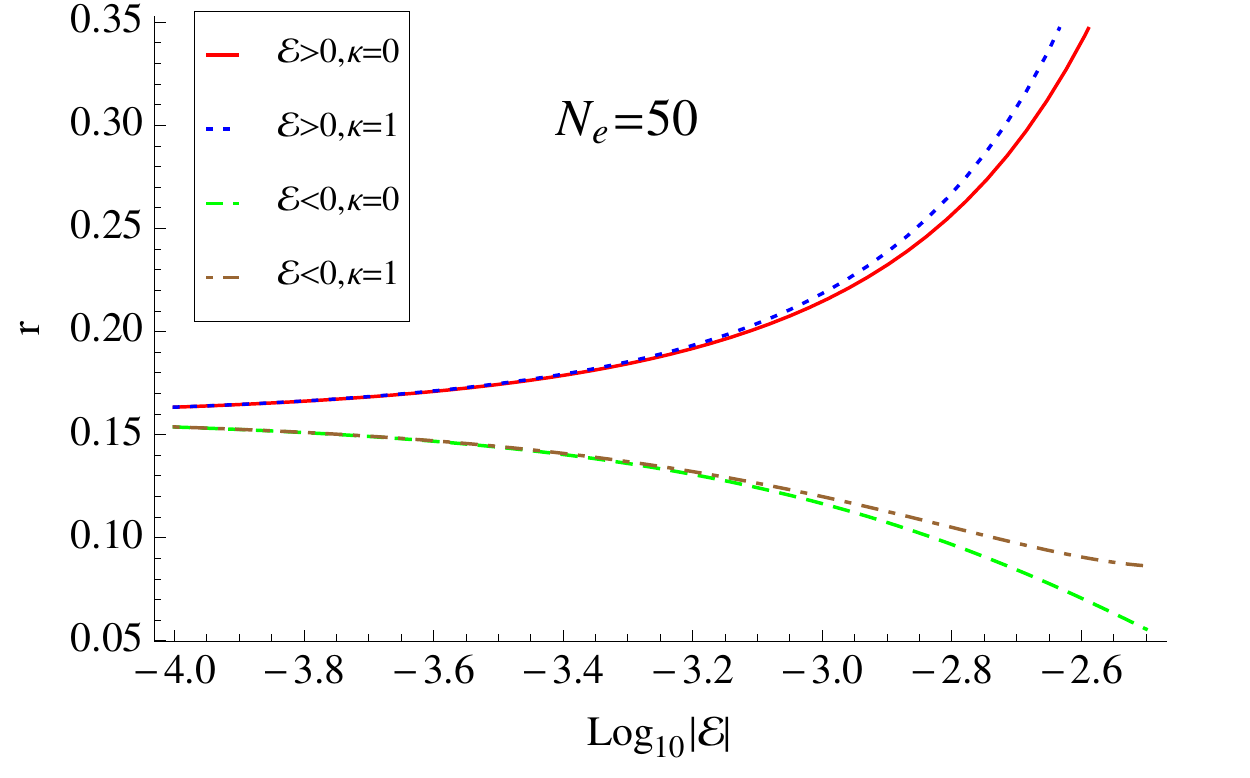}
 \end{center}
\caption{\sl \small
The tensor fration $r$ as a function of the shift-symmetry breaking parameter ${\cal E}$ for $\kappa=0,1$.
}
\label{fig:fraction}
\end{figure}

The spectral index $n_s$ and the tensor fraction $r$ are given by
\begin{eqnarray}
n_s = 1-6\epsilon + 2\eta,~~
r = 16\epsilon.
\end{eqnarray}
In Figs.~\ref{fig:tilt} and \ref{fig:fraction}, we show $n_s$ and $r$ as functions of ${\cal E}$ for $\kappa=0$ and $1$.
It can be seen that $n_s$ and $r$ are significantly altered by the shift symmetry breaking expressed by ${\cal E}$.

\begin{figure}[tb]
\begin{center}
  \includegraphics[width=.6\linewidth]{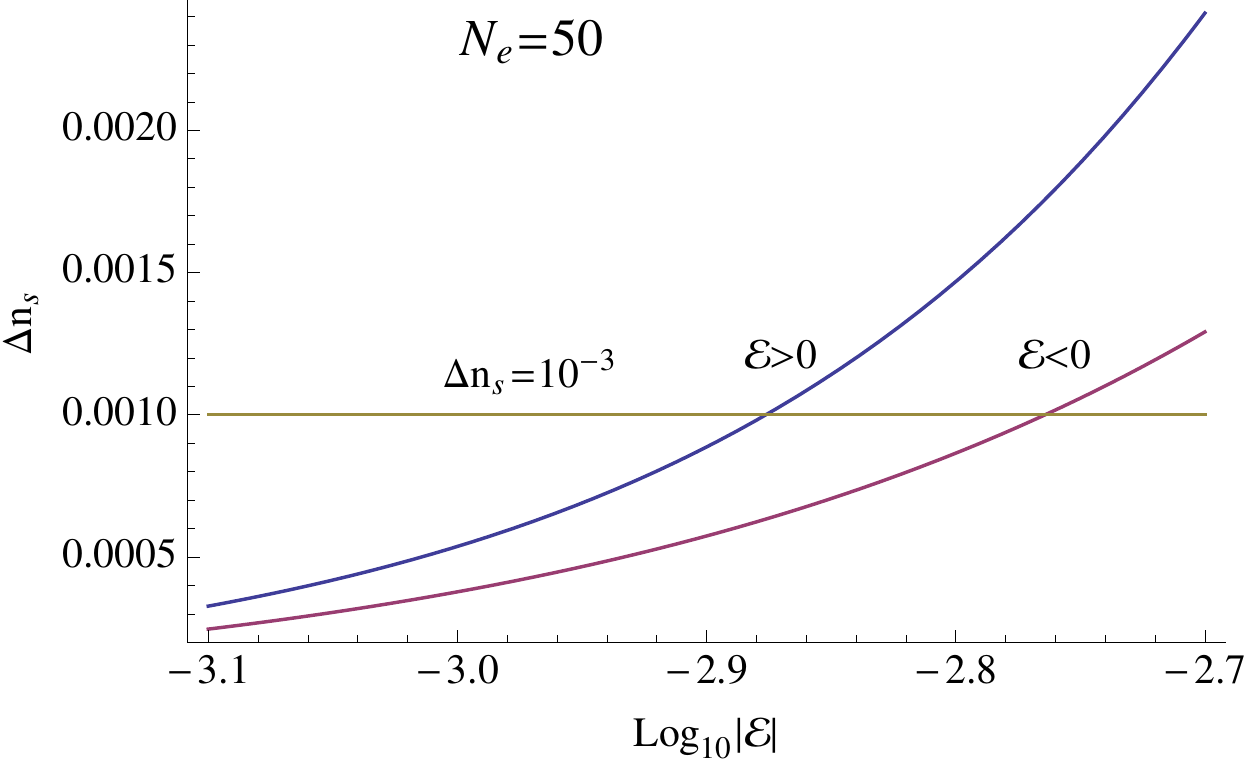}
 \end{center}
\caption{\sl \small
Sensitivity of the spectral index $n_s$ to higher dimensional terms in the Kahler potential.}
\label{fig:tilt-d}
\end{figure}

\begin{figure}[tb]
\begin{center}
  \includegraphics[width=.6\linewidth]{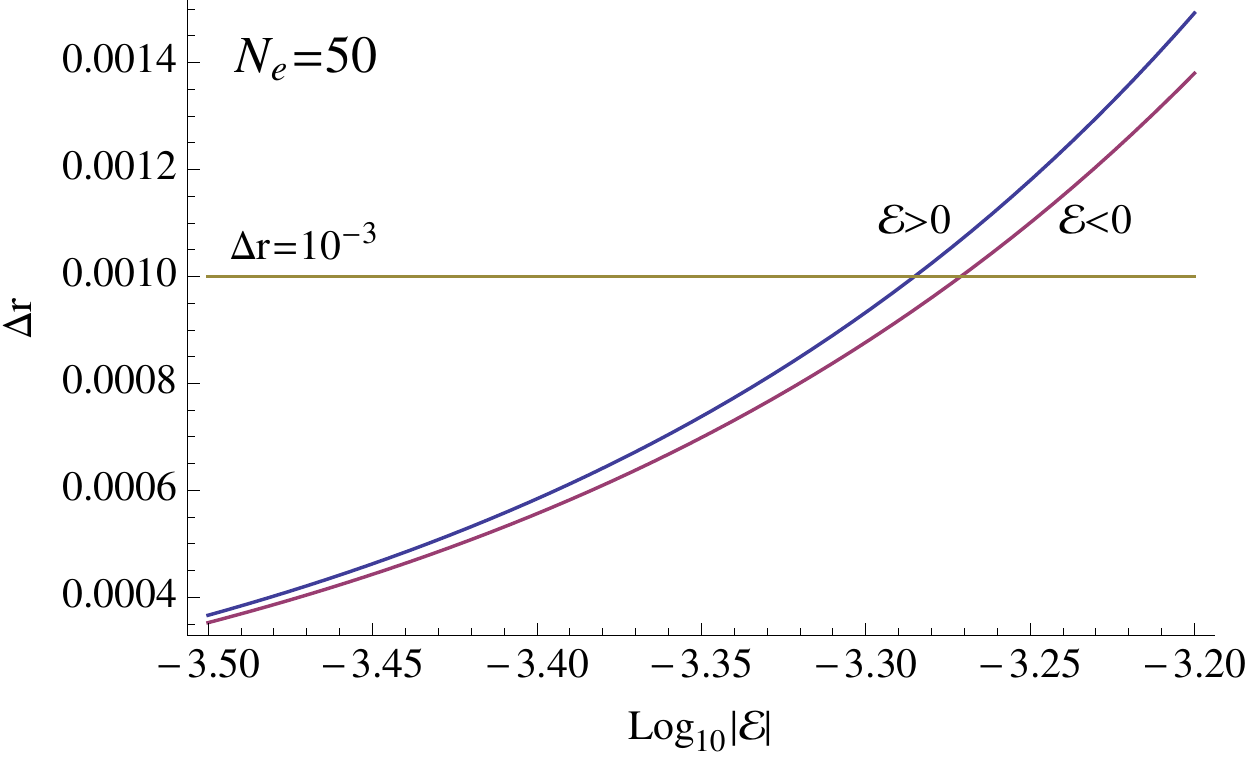}
 \end{center}
\caption{\sl \small
Sensitivity of the tensor fraction $r$ to higher dimensional terms in the Kahler potential.}
\label{fig:fraction-d}
\end{figure}

Let us estimate the bound on ${\cal E}$ such that higher dimensional ${\cal O}({\cal E}^2)$ terms in the Kahler potential are negligible.
Since we have expressed the magnitude of ${\cal O}({\cal E}^2)$ terms by the parameter $\kappa$ (see Eq.~(\ref{eq:Kahler})), we can estimate the bound on 
${\cal E}$ by investigating the dependence of the prediction for $n_s$ and $r$ on $\kappa$.
In Figs.~\ref{fig:tilt-d} and \ref{fig:fraction-d}, we show $\Delta n_s\equiv |n_{s,\kappa=1}-n_{s,\kappa=0}|$ and $\Delta r\equiv|r_{s,\kappa=1}-r_{s,\kappa=0}|$ as functions of ${\cal E}$.
We define the insensitivity to higher dimensional ${\cal O}({\cal E}^2)$ terms as $\Delta n_s < 10^{-3}$ and $\Delta r < 10^{-3}$, which is the typical
resolution of future satellite experiments such as the CMBPol~\cite{Baumann:2008aq} and the LiteBIRD~\cite{LiteBIRD}.
From Figs.~\ref{fig:tilt-d} and \ref{fig:fraction-d}, we put a bound on ${\cal E}$ as
\begin{eqnarray}
\label{eq:bound}
|{\cal E}|< 10^{-3.3}.
\end{eqnarray}

\begin{figure}[tb]
\begin{center}
  \includegraphics[width=.8\linewidth]{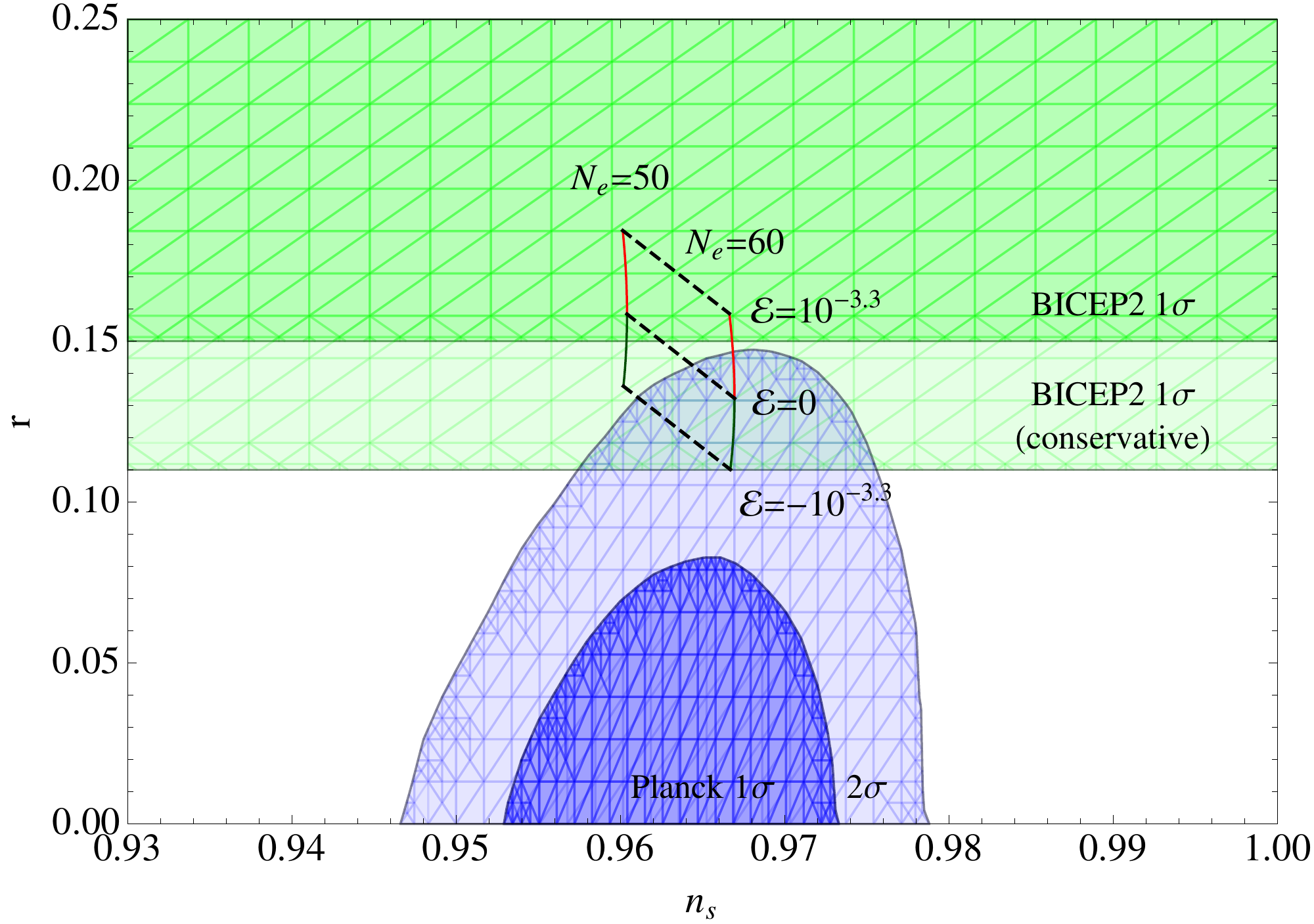}
 \end{center}
\caption{\sl \small
The prediction on the spectral index $n_s$ and the tensor fraction $r$ for the $Z_2$ symmetric model.
We also show the constraint from the Planck and the BICEP2 experiments.
}
\label{fig:tilt-fraction}
\end{figure}

In Fig.~\ref{fig:tilt-fraction}, we show the prediction on $n_s$ and $r$ for $|{\cal E}|<10^{-3.3}$.
We also show constraints from the Planck experiment~\cite{Ade:2013uln} and the BICEP2 experiment~\cite{Ade:2014xna}
for the pivot scale of $0.002~{\rm Mpc}^{-1}$.
%It can be seen that the chaotic inflation model is consistent with the constraints within $1\sigma$ level.
It can be seen that the prediction of the model can lie between the results of the Planck and the BICEP2 experiments (see also Ref.~\cite{Kallosh:2010ug}).
We stress that the prediction is not affected by higher dimensional terms in the Kahler potential as long as the constraint given in Eq.~(\ref{eq:bound}) is satisfied.

Note that $n_s$ and $r$ also depend on $N_e$.
Difference of the reheating temperature by an order of magnitude
%which is of importance in considering the thermal history of the universe,
changes $N_e$ corresponding to the pivot scale by ${\cal O}(1)$. The ${\cal O}(1)$ change in $N_e$ also modifies the prediction on $n_s$ and $r$ by ${\cal O}(10^{-3})$.
Therefore, by measuring $n_s$ and $r$ with an accuracy of ${\cal O}(10^{-3})$, we can quantify the reheating temperature of the universe
%~\cite{Nakayama:2008wy}
within a factor of ${\cal O}(10)$.
It should be noted that this is possible only within the parameter range given in Eq.~(\ref{eq:bound}).

In the above analysis, we have concentrated on the quadratic chaotic inflation model.
This is because the constraint on the spectral index $n_s\simeq 0.96$ favors the quadratic model.
In Ref.~\cite{Spergel:2013rxa}, however, it is pointed out that the central value of $n_s$ is larger and amounts to $\simeq 0.97$.
If that is the case,
models with lower power potentials~\cite{Takahashi:2010ky,Harigaya:2012pg} are favored.
For models with lower power potentials, we can discuss the effect of non-holomorphic shift-symmetry breaking in the similar way as we have done in this letter.

\section{Discussion and conclusions}
In this letter, we have reconsidered chaotic inflation models in the SUGRA.
We have introduced a non-holomorphic shift-symmetry breaking parameter ${\cal E}$ and discussed its effect on the inflaton dynamics.
We have clarified the range of ${\cal E}$ where higher dimensional terms are negligible for the inflaton dynamics and the model possesses predictability and testability.
We have shown that the prediction for the spectral index $n_s$ and the tensor fraction $r$ are
given by $n_s\sim 0.96$ and $r = 0.11 - 0.18$.
% with a small shift-symmetry breaking parameter.
The prediction can lie between the results of the Planck and the BICEP2 experiments.
It is interesting that future experiments will measure $n_s$ and $r$ accurately and reveal the structure of the shift-symmetry breaking in the inflaton sector.
We have also shown that future observations of the CMB can quantify the reheating temperature of the universe within a factor of ${\cal O}(10)$, as long as ${\cal E}$ is in the range we have clarified.

%We have shown that the chaotic inflation model is consistent with observations with sufficiently small shift symmetry breaking in the Kahler potential, so that higher dimensional terms in the Kahler potential are negligible for the inflaton dynamics, and hence the model possesses predictability.

Let us comment on the magnitude of the shift symmetry breaking.
We have introduced two shift-symmetry breaking parameters, $m$ and ${\cal E}$.
The magnitude of the curvature perturbation indicates that $m\sim 10^{-5}$ and the consistency with the observed spectral index and the tensor fraction suggests that $|{\cal E}| \sim 10^{-3}$.
Therefore, the two shift-symmetry breaking parameters are different by order of magnitudes.
Note that the $m$ is a holomorphic parameter while ${\cal E}$ is a non-holomorphic one,
and hence they may have different origins.
We hope that a more fundamental theory explains the origin of the shift-symmetry breaking.

\begin{figure}[tb]
\begin{center}
  \includegraphics[width=.8\linewidth]{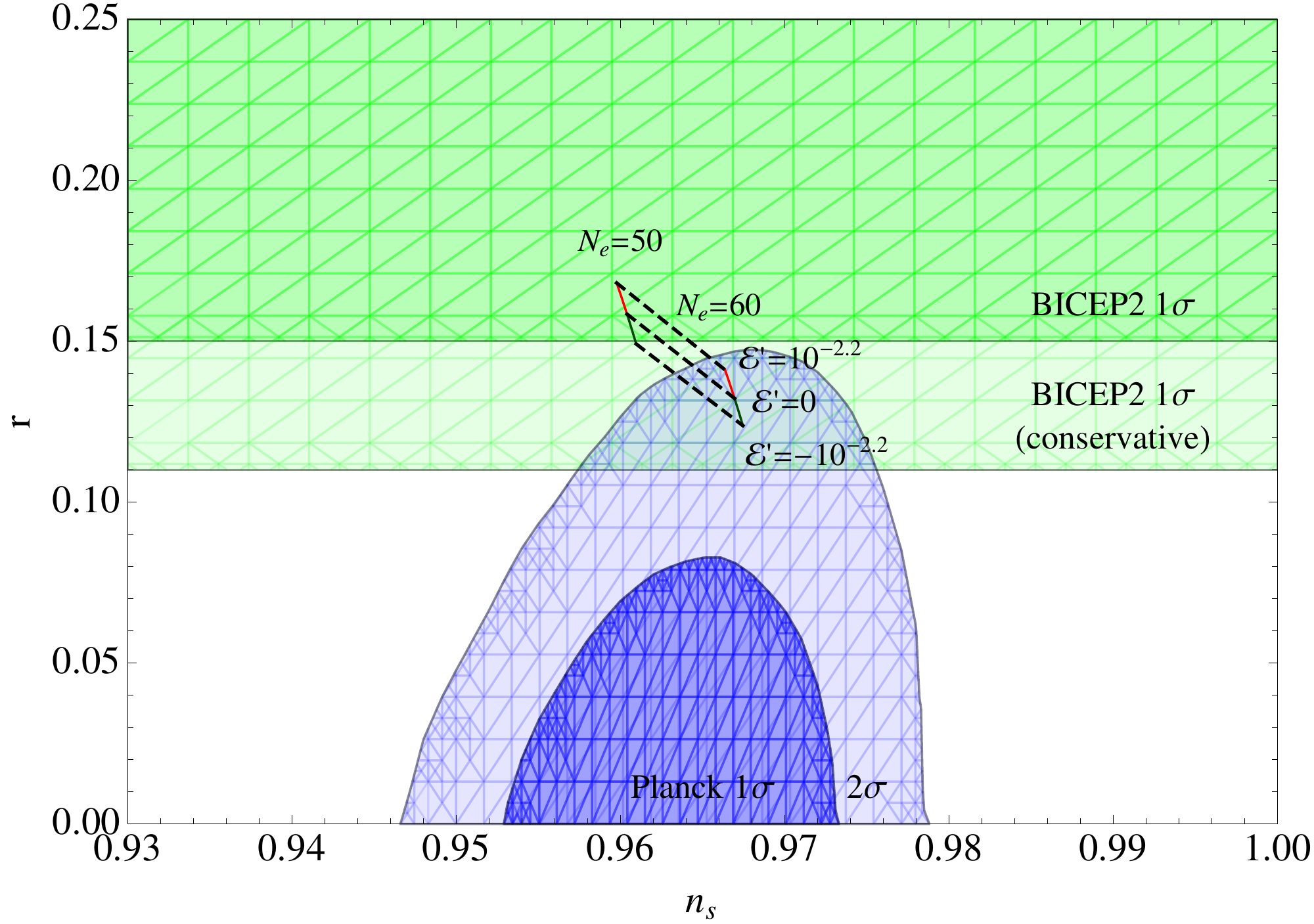}
 \end{center}
\caption{\sl \small
The prediction on the spectral index $n_s$ and the tensor fraction $r$ for the model without the $Z_2$ symmetry.
We also show the constraint from the Planck and the BICEP2 experiments.
}
\label{fig:tilt-fraction-2}
\end{figure}

Finally, let us briefly consider a model without the $Z_2$ symmetry. In this case, the Kahler potential is expanded as
\begin{eqnarray}
K = c (\Phi + \Phi^*) + \frac{1}{2} (\Phi + \Phi^*)^2 - i \frac{{\cal E}'}{\sqrt{2}} (\Phi - \Phi^*) - \frac{\kappa'}{2}\left(\frac{{\cal E}'}{\sqrt{2}}\right)^2 (\Phi - \Phi^*)^2  + \cdots,
\end{eqnarray}
and the scalar potential of the inflaton is given by
\begin{eqnarray}
V (\phi) = {\rm exp}\left( {\cal E}' \phi + \frac{\kappa'}{2}{\cal E}'^2 \phi^2 + \cdots \right) \frac{1}{2} m^2 \phi^2.
\end{eqnarray}
We can clarify the predictability of the model, that is, insensitivity to $\kappa'$, as we have done in this letter.
It can be shown that the model possesses the predictability as long as
\begin{eqnarray}
|{\cal E}'| < 10^{-2.2}.
\end{eqnarray}
The prediction on $n_s$ and $r$ for $|{\cal E}'| < 10^{-2.2}$ is shown in Figure~\ref{fig:tilt-fraction-2}.
Here, we have assumed that the inflaton field value is positive during the inflation.

\section*{Acknowledgments}
This work is supported by Grant-in-Aid for Scientific research from
the Ministry of Education, Culture, Sports, Science, and Technology (MEXT), Japan, No.\ 22244021 (T.T.Y.),
and also by World Premier International Research Center Initiative (WPI Initiative), MEXT, Japan.
 The work of K.H. is supported in part by a JSPS Research Fellowships for Young Scientists.

\end{document}